\documentclass[
reprint,
superscriptaddress,
amsmath,
amssymb,
aps,
longbibliography,
prx
]{revtex4-2}
\usepackage{amsmath}
\usepackage{mathtools}
\usepackage{amssymb}
\usepackage{graphicx}
\usepackage{bm}
\usepackage[colorlinks, linkcolor=blue, citecolor=blue, urlcolor=blue,breaklinks=true]{hyperref}
\usepackage{color,dsfont,multirow,booktabs}
\usepackage{braket}

\begin{document}

\title{Quantum enhanced probing of multilayered-samples}

\author{Mayte Y. Li-Gomez}
\altaffiliation{These authors contributed equally to this work.}
\affiliation{Institute for Quantum Science and Technology, and Department of Physics and Astronomy University of Calgary,
2500 University Drive NW, Calgary, Alberta T2N 1N4, Canada}
\author{Pablo Yepiz-Graciano}
\altaffiliation{These authors contributed equally to this work.}
\affiliation{Instituto de Ciencias
Nucleares, Universidad Nacional Autonoma de México,
04510 Ciudad de México, México}
\author{Taras Hrushevskyi}
\altaffiliation{These authors contributed equally to this work.}
\affiliation{Institute for Quantum Science and Technology, and Department of Physics and Astronomy University of Calgary,
2500 University Drive NW, Calgary, Alberta T2N 1N4, Canada}
\author{Omar Calderon-Losada}
\affiliation{Centro de Investigación e Innovación en Bioinformática y Fotónica (CIBioFi). Departamento de Física, Universidad del Valle, Calle 13 No. 100-00, 760032, Cali, Colombia}
\author{Erhan Saglamyurek}
\affiliation{Institute for Quantum Science and Technology, and Department of Physics and Astronomy University of Calgary,
2500 University Drive NW, Calgary, Alberta T2N 1N4, Canada}
\author{Dorilian Lopez-Mago}
\affiliation{Tecnológico de Monterrey, Escuela de Ingenier\'{i}a y Ciencias, Monterrey, N.L. 64849, México}
\author{Vahid Salari}
\affiliation{Institute for Quantum Science and Technology, and Department of Physics and Astronomy University of Calgary,
2500 University Drive NW, Calgary, Alberta T2N 1N4, Canada}
\author{Trong Ngo}
\affiliation{Institute for Quantum Science and Technology, and Department of Physics and Astronomy University of Calgary,
2500 University Drive NW, Calgary, Alberta T2N 1N4, Canada}
\author{Alfred B. U'Ren}
\affiliation{Instituto de Ciencias
Nucleares, Universidad Nacional Autonoma de México,
04510 Ciudad de México, México}
\author{Shabir Barzanjeh}
\email{shabir.barzanjeh@ucalgary.ca}
\affiliation{Institute for Quantum Science and Technology, and Department of Physics and Astronomy University of Calgary,
2500 University Drive NW, Calgary, Alberta T2N 1N4, Canada}
\date{\today}

\begin{abstract}
Quantum sensing exploits quantum phenomena to enhance the detection and estimation of classical parameters of physical systems and biological entities, particularly so as to overcome the inefficiencies of its classical counterparts. A particularly promising approach within quantum sensing is Quantum Optical Coherence Tomography which relies on non-classical light sources to reconstruct the internal structure of multilayered materials. Compared to traditional classical probing, Quantum Optical Coherence Tomography provides enhanced-resolution images and is unaffected by even-order 
dispersion. One of the main limitations of this technique lies in the appearance of artifacts and echoes, i.e. fake structures that appear in the coincidence interferogram, which hinder the retrieval of information required for
tomography scans. Here, by utilizing a full theoretical model, in combination with a fast genetic algorithm to post-process the data, we successfully extract the morphology of complex multilayered samples and thoroughly
distinguish real interfaces, artifacts, and echoes. We test the effectiveness of the model and algorithm by comparing its predictions to  experimentally-generated interferograms through the controlled variation of the pump wavelength. Our results could potentially lead to the development of practical high-resolution probing of complex structures and non-invasive scanning of photo-degradable materials for biomedical imaging/sensing, clinical applications, and materials science.
\end{abstract}
\maketitle
\section{Introduction}
Quantum sensing is one of the emerging technologies that exploits quantum correlations and entanglement for detecting and imaging classical objects with fundamentally better performance while providing opportunities to overcome the limitations and challenges of the traditional classical approaches \cite{RMP, Rev2, PhysRevLett.114.080503, Pirandola2018, radar, rev3, Giovannetti2011, PhysRevA.107.032611, Salari2023}. One of the well-developed and powerful methods of quantum sensing is Quantum Optical Coherence Tomography (QOCT) that utilizes the unique features of entangled photon pairs to image different layers within a sample of interest \cite{Abouraddy_2002, NASR20091154, Teich2012, Lopez-Mago:12, Lavoie:09, Yepiz_2019, ebi}. This quantum imaging technique offers unprecedented capabilities in biomedical imaging \cite{nasr_2009, yepiz-borja_2022} and full-field scanning of samples using cameras with a single-photon sensitivity \cite{Ibarra-Borja_2020,Bienvenu_ndagano_2022}. Compared to its classical counterpart, QOCT offers higher image (axial) resolution by a factor of 2 and provides immunity to even-order dispersion effects, the latter ensured by strict frequency anti-correlation.

Nonetheless, the vast practicality of QOCT is often hampered by long acquisition times and the problem of artifacts, i.e. fake structures appearing as peaks or dips in the coincidence interferograms that corrupt the QOCT signal and hinder the proper interpretation of tomography scans. Such fake structures are inherent to the system due to the quantum nature of QOCT. Note also that this difficulty becomes increasingly acute as the sample complexity, in terms of the number of interfaces, increases. For even a modest number of interfaces, peak/dip discrimination can become an extremely challenging task. Previous works have developed tools geared at correctly inferring the sample morphology from the QOCT interferogram through artifact removal. In Ref.~\cite{Yepiz_2019} a specific method was demonstrated for suppressing cross-interference artifacts based on Spontaneous Parametric Down Conversion (SPDC) photon pairs produced by a broadband, femtosecond-duration pump.   At the core of this technique is the existence of multiple pump wavelengths, which leads to the possibility of spectrally sweeping a continuous-wave pump and averaging the resulting interferograms as an alternative method.   It has  been shown that artifact removal can likewise be attained through the Fourier transformation properties of the two-dimensional joint spectrum~\cite{Kolenderska:20, Kolenderska_2021}, and using deep-learning techniques~\cite{Liu_22}.
Both artifact removal approaches are based on a theoretical framework that is limited to a model based on only two interfaces and which requires processes that are prone to loss of relevant information. Moreover, such models have disregarded the presence of additional structures appearing in the QOCT interferogram which, though neglected, have a large impact on the signal as a whole.

In this work, rather than focusing directly on artifact removal, we aim first to develop a QOCT model which fully takes into account all observed effects, some of which have been neglected in past work.   This includes multiple reflections in the sample, leading to \emph{echoes} which behave as 
virtual interfaces, as well as cross-interference artifacts between a real interface and an echo, and even cross-interference artifacts between two echoes.  As will become clear below, on the one hand these effects can be significant for realistic situations, and on the other hand, multiple terms can contribute to a given interferogram peak/dip so that attempting to infer the sample morphology without an adequate physical understanding will yield misleading results. Armed with a full understanding of the relevant optical effects that occur in the sample, in this work we aim secondly at developing a computational tool, based on genetic algorithms, that can predict the morphology which best fits a particular observed QOCT interferogram. Through this approach, we are particularly able to extract sample parameters (layer optical thicknesses and interface reflectivities) from the QOCT interferogram. Our methodology will help to accurately interpret the often highly-complex interferograms which result from realistic samples and consequently pave the way toward the practical implementation of quantum scanning and probing. 

\section{Theoretical modeling}

\subsection{Multi-layer sample description and QOCT signal} \label{sec:theo}
\begin{figure}
    \centering
    \includegraphics[width=\columnwidth]{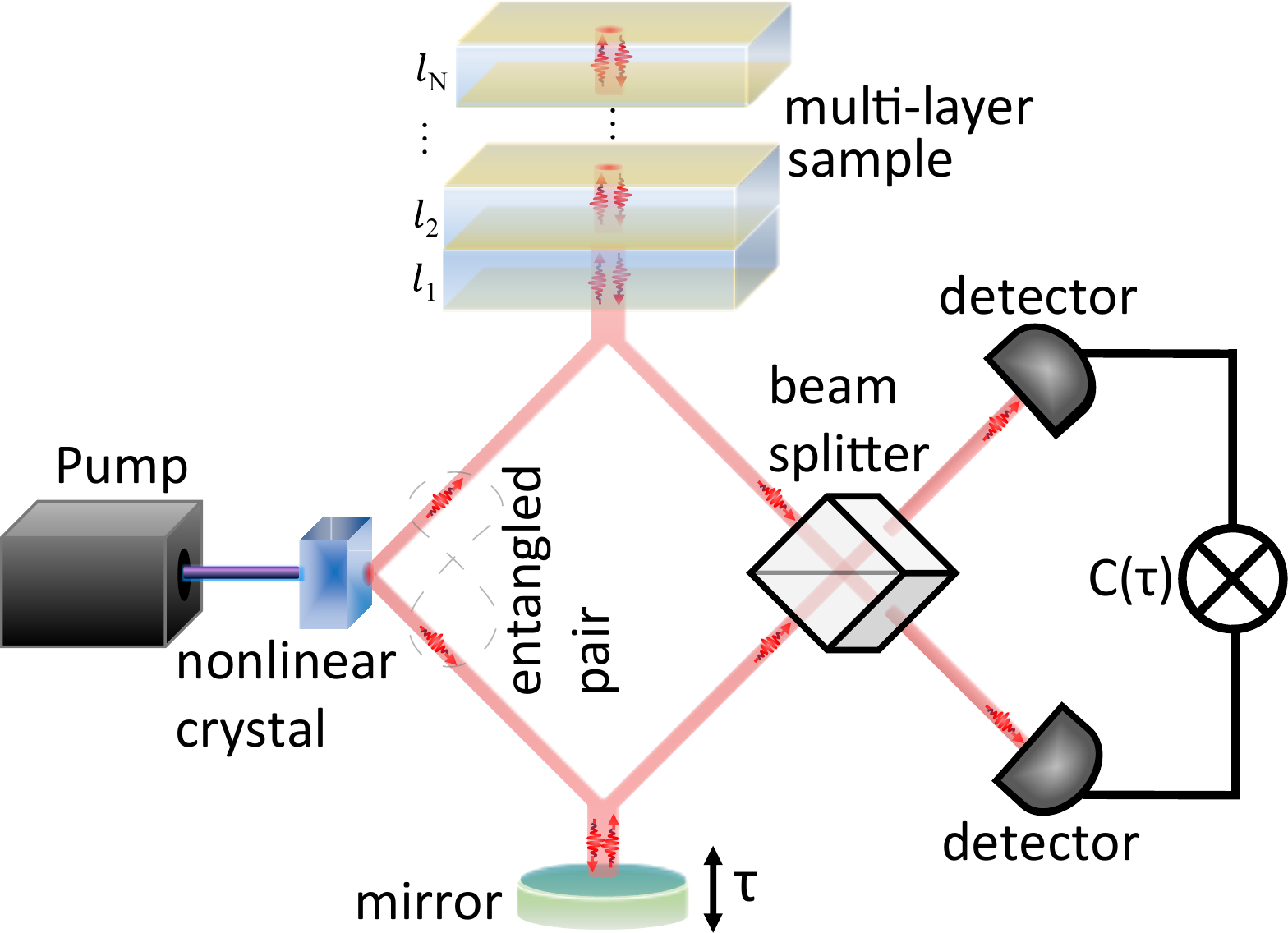}
    \caption{\textbf{Schematic of the Setup}: The entangled idler-signal photon pairs are generated through the SPDC process, by a pump that impinges on a nonlinear crystal. The idler photons are used to probe different layers of an unknown sample while the signal photons are transmitted through a reference arm of known tunable length to match the path delay $\tau$ with the idler photons. The signal and idler photons are then recombined at a beamsplitter where Hong-Ou-Mandel interference occurs. Measuring the coincidence rate signal $C(\tau)$ leads to an interferogram that carries information about the internal interfaces in the sample. }
    \label{fig:setup_scheme}
\end{figure}

As shown in Fig. \ref{fig:setup_scheme}, QOCT is generally based on Hong-Ou-Mandel (HOM) \cite{ebi} interferometry in which  pairs of entangled photons, typically generated through SPDC, are split and conducted through two different paths. The signal photon travels through a reference path of known length, while the idler photon passes through a sample arm which reflects light back from each of the multiple interfaces in a sample. The signal and idler photons are then recombined at a beamsplitter where HOM interference occurs and, provided that the two photons are indistinguishable,  will exit the same port in a process referred to as bunching. Photon-bunching is more likely to happen when the temporal delay between the signal and idler photons is zero ($\tau =0$); that is, when the path lengths of the reference arm and the sample arm are equal. The HOM interference is characterized by a drop in the coincidence count rate, or HOM dip, of the output joint detection at $\tau =0$.  Thus, for every interface, a corresponding HOM dip will appear in the coincidence rate signal $C(\tau)$, resulting in a sequence of dips in the interferogram obtained from a multilayer sample. 

In order to describe the  QOCT output signal, we consider a non-conservative system consisting of a sample in the form of a dielectric
multilayer stack, interleaved with metallic layers, as shown in Fig. \ref{fig:setup_scheme}.  The sample is conveniently described as the concatenation of the wave-transfer matrices representing light with wavelength $\lambda_0$ impinging on each dielectric boundary, described by \cite{saleh_teich2007}

\begin{align}
\label{eq:M_boundary_and_prop}
\mathbf{M}_{k,l} & =\frac{1}{t_{lk}} 
\begin{bmatrix}
t_{kl}t_{lk}-r_{kl}r_{lk} & r_{lk}\\
-r_{kl} & 1
\end{bmatrix}, 
\end{align}

and light transmitted between boundaries $\mathbf{M}_k=\mathbf{M}_k^{\varphi} \mathbf{M}_k^{\kappa}$
\begin{align}
\mathbf{M}_k^{\varphi} & = 
\begin{bmatrix}
e^{-i\varphi_k} & 0\\
0 & e^{i\varphi_k}
\end{bmatrix}, \label{eq:propagationHM}\\
    \mathbf{M}_k^{\kappa}&=\begin{bmatrix}
    e^{-\kappa} & 0\\
0 & e^{\kappa}
    \end{bmatrix},
\end{align}
$|t_{kl}|^2+|r_{kl}|^2=1$ 
where the complex quantities $t_{kl}$ and $r_{kl}$ are the amplitude transmissivity and reflectivity coefficients, for a wave incident from medium $k$ towards medium $l$.  Our approach is to assume that each interface is lossless, while separately including a loss matrix, so that the following four constraints are fulfilled: $|r_{kl}|=|r_{lk}|$,  $|t_{kl}|=|t_{lk}|$,  $|t_{kl}|^2+|r_{kl}|^2=1$,  and $t_{lk}/t_{kl}*=- r_{lk}/r_{kl}*$.
For each  layer $k$, the phase argument $\varphi_k = k_0 n_k d_k$ is defined in terms of the free-space wavenumber $k_0=2\pi/\lambda_0$, the width $d_k$, and the refractive index $n_k$. 

We also consider the presence of lossy layers, such as a metallic coating prior to medium $k$, in the wave-transfer matrix by introducing $\kappa=\alpha L_k$ where $\alpha$ is the absorption coefficient and $L_k$ is the width of the lossy layer. We write the matrix describing propagation, $\mathbf{M}_k^{\varphi}$, and the matrix accounting for losses, $\mathbf{M}_k^{\kappa}$, as separate matrices in order to generalize the model to artificial media, where metallic and non-metallic interfaces could be interleaved in different ways.

The wave-transfer matrix of a sample composed of $N$ layers (i.e., $N+1$ interfaces) is written as

\begin{equation}\label{eq:overallWTM}
\mathbf{M} = \prod_{j=1}^{N \leftarrow} \mathbf{M}^\varphi_{j}\mathbf{M}^\kappa_{j} \mathbf{M}_{j-1,j} = \begin{bmatrix}
A & B\\
C & D
\end{bmatrix}
\end{equation}

\noindent where the left arrow in the upper limit indicates that the $(j+1)$th factor precedes the $j$th factor in the order of multiplication.

The scattering matrix $\mathbf{S}$ can now be calculated from the overall transfer matrix $\mathbf{M}$ \cite{saleh_teich2007} written in terms of overall forward  ($r_{sample}$ and $t_{sample}$) and overall backward ($r'_{sample}$ and $t'_{sample}$) reflection/transmission coefficients 

\begin{equation}\label{eq:scatteringMatrix}
\mathbf{S} = \frac{1}{D} \begin{bmatrix}
AD-BC & B\\
-C & 1
\end{bmatrix} =\begin{bmatrix}
t_{sample} & r'_{sample}\\
r_{sample} & t'_{sample}
\end{bmatrix}.
\end{equation}

Since the QOCT measurement is based on the light reflected from the sample, we use the effective reflection coefficient given by $r_{sample}=-C/D$. As a consequence, the multi-layered sample can be characterized by the sample transfer function given by
\begin{equation}\label{eq:srt}
    H(\omega) = -\frac{C(\omega)}{D(\omega)},
\end{equation}
where the frequency dependence in the above equation comes from the phase term $\varphi_k = k_0(\omega) n_k(\omega) d_k$. For materials that are not  highly dispersive, i.e. which fulfill 
$\frac{dn(\omega)}{d\omega} \ll  \frac{n}{\omega}$, this phase may be approximated as $\varphi_k=\omega \tau'_k$ where $\tau'_k$ represents the optical thickness in time units.

In order to gain physical insight regarding the effects of multiple reflections, we begin with the simple case in which light impinges on a single layer (i.e. two interfaces), surrounded by air.  Such a sample can be described by the amplitude transmissivities $t_{01}$ and $t_{10}$ ($t_{12}$ and $t_{21}$), and the reflectivities $r_{01}$ and $r_{10}$ ($r_{12}$ and $r_{21}$), for the first (second) interface. From Eq.~(\ref{eq:overallWTM}), the overall wave-transfer matrix is given by $\mathbf{M}=\mathbf{M}_{1,2}\mathbf{M}^\varphi_{1}\mathbf{M}^\kappa_{1}\mathbf{M}_{0,1}$. The  sample transfer function obtained after using Eq.~(\ref{eq:scatteringMatrix}) and Eq.~(\ref{eq:srt}) is
\begin{align}\label{eq:transfer_function_1layer}
    H(\omega)&=r_{01}+\frac{r_{12}t_{01}t_{10}e^{-2(\kappa+i\varphi)}}{1-r_{10}r_{12}e^{-2(\kappa+i\varphi)}}\\ 
    &= r_{01} + r_{12}t_{01}t_{10} e^{-i\omega T} + r_{10}r_{12}^2t_{01}t_{10} e^{-i\omega (2T)}+...,\nonumber
\end{align}
where we have defined $T\equiv 2\tau'$, which represents the round trip from interface 0 to interface 2, and back to interface 0, and where we have renamed $r_{12} e^{-2\kappa} \rightarrow r_{12}$, thus lumping the metallic-layer loss at the first interface into the second-interface reflectivity.  Note that this result does not depend on the second interface transmissivities $t_{12}$ and $t_{21}$, and is likewise unaffected by losses due to a metallic layer on the second interface, since the QOCT signal excludes any light transmitted by the sample as a whole (in general, for $n$ interfaces, losses at the $n$th interface can be ignored).  


Let us assume that the reference arm is balanced with respect to light reflected in the sample arm  from the first interface.  Under these circumstances, the first term corresponds to light directly reflected from the first interface and occurs at zero delay ($\tau=0$).
The second term, which occurs at a delay $\tau=T$ is reduced by a factor $r_{12}t_{01}t_{10}$ due to two passes through the first interface.
The third term, which occurs at a delay $\tau=2T$, represents the case in which the signal photon travels from interface 1 to interface 2 and back, and then once more from interface 1 to interface 2 and back, so that it traverses the layer four times, attenuated by a factor $r_{10}r_{12}^2t_{01}t_{10}$.
This corresponds to the lowest-order multiple reflection, and leads to a term that behaves as a virtual interface at twice the distance from interface 1 as the separation between the two interfaces.
We refer to such signals due to multiple reflections as \emph{echoes}.
Note that in addition to the artifacts already discussed (cross interference between interfaces), echo-induced artifacts can originate from cross interference between an echo and an interface, as well as between two echoes.

Note that echoes and echo-induced artifacts become a source of additional complexity in the resulting interferogram, especially since as will be discussed below multiple terms can co-exist at a given delay value.
This complicates the task of inferring the sample morphology from the QOCT interferogram.   
The next section shows examples of QOCT interferograms and our strategy to correctly identify all terms: interfaces, artifacts, echoes, and echo-induced artifacts.


\subsection{Coincidence rate }

In this section, we develop a full theoretical model of QOCT using our transfer-matrix  approach and find an expression for the interferogram and its dependence on the sample parameters, associated with the different layers and interfaces. 
Note that while here we obtain analytic expressions for the 1-layer (2-interface) case,  the same approach can be utilized numerically for $\ge 3$ interfaces and also note that the expressions provide insights that are valid for arbitrary samples.
The quantum state of entangled photon pairs generated through the process of SPDC in a $\chi^{(2)}$ nonlinear crystal can be expressed as \cite{2007}
\begin{equation}
\ket{\psi}=\ket{0}+\eta \int \int d\omega_1 d\omega_2 f(\omega_1,\omega_2) \ket{\omega_1}\ket{\omega_2},
\end{equation}
where $f(\omega_1,\omega_2)$ is the joint spectral amplitude determined by the phase-matching conditions, and where $\eta$ accounts for the SPDC process efficiency. The idler and signal photons are described by $\ket{\omega_i}$ with $i=1,2$.

An expression for the QOCT interferogram follows from modeling the coincidence count rate at the interferometer output as a function of the temporal delay $\tau$ between the idler photon which traverses the sample, and the signal photon which traverses the reference arm.  It is given by \cite{Yepiz_2019}
\begin{align}
\label{eq:coincidences}
C(\tau)&=\frac{N_0}{4} \int \int d\omega_1 d\omega_2 I(\omega_1,\omega_2) \nonumber\\  &=\Gamma_0-2\mathrm{Re}\{\Gamma(\tau)\},
\end{align}
where $N_0$ is the background coincidence count level and
\begin{multline*}
    I(\omega_1,\omega_2)=\\
    \left \vert f(\omega_1,\omega_2)H(\omega_2)-f(\omega_2,\omega_1)H(\omega_1)e^{i(\omega_2-\omega_1) \tau} \right\vert^2, 
\end{multline*} in which $f(\omega_1,\omega_2)$ is the joint spectral amplitude, assumed henceforth to be symmetric, i.e. $f(\omega_1,\omega_2)=f(\omega_2,\omega_1)$.
The constant term $\Gamma_0$ represents the coincidence count rate at large delay $\vert\tau\vert \gg T$
\begin{multline}
\Gamma_0=\frac{N_0}{4} \int \int d\omega_1 d\omega_2 S(\omega_1,\omega_2)\\ \times \left[ \left \vert H(\omega_1) \right\vert^2 +\left \vert H(\omega_2) \right\vert^2 \right],
\end{multline}
and
\begin{multline}
\label{eq:cross-interference}
\Gamma(\tau)=\frac{N_0}{4} \int \int d\omega_1 d\omega_2 S(\omega_1,\omega_2) \\ \times H(\omega_2) H^*(\omega_1)e^{i(\omega_2-\omega_1)\tau}
\end{multline}
shows the interference term,  where  $S(\omega_1,\omega_2)\equiv|f(\omega_1,\omega_2)|^2$ is the joint spectral intensity  which can be approximated  as a Gaussian function~\cite{Yepiz_2019}

\begin{figure}[t]
    \centering
    \includegraphics[width=\columnwidth]{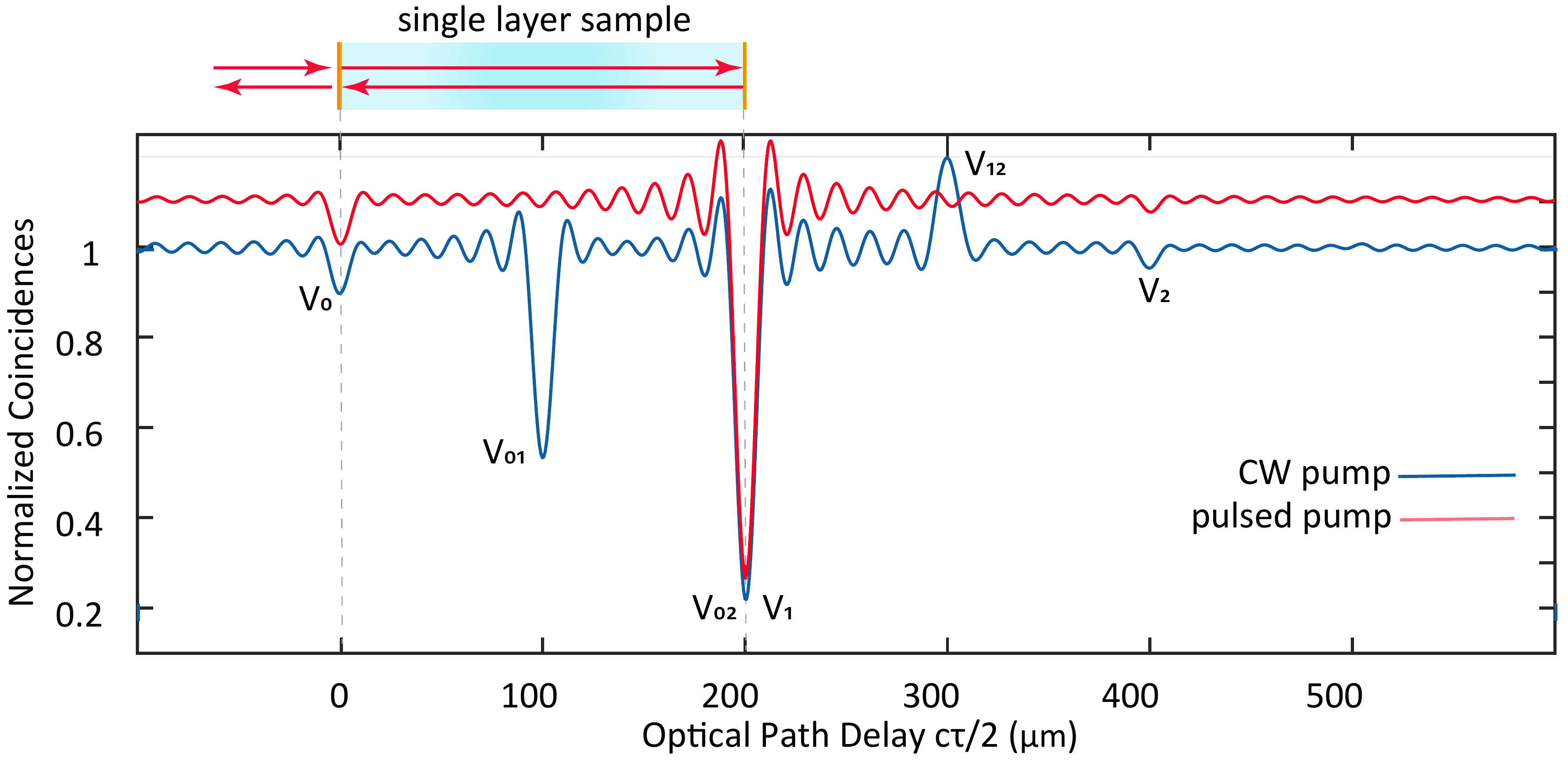}
    \caption{Single-layer (two-interface) interferogram corresponding to the first three terms of the transfer function in the Eq. (\ref{eq:transfer_function_1layer}), where we have assumed $cT=200\mu$m. Here we compare the result of QOCT for a monochromatic CW pump  (blue line) with a pulsed (femtosecond) pump (red line). The interferogram obtained in the CW case shows the two dips associated with the  boundaries $V_0$, $V_1$, the first echo $V_2$ and three cross-terms $V_{01}$, $V_{02}$, $V_{12}$, while in the pulsed pump case artifacts are suppressed. The red line has been vertically displaced for graphical clarity.  Note that sinc-type oscillations arise from a rectangular bandpass filter acting on the signal and idler photons.}
    \label{fig:two_boundaries_artifacts}
\end{figure}
\begin{eqnarray}
\label{eq_joint_spectral_intensity}
S(\omega_1,\omega_2)=\frac{4}{\pi \Omega_a \Omega_d} 
&\exp&\left[-2\left(\frac{\omega_1+\omega_2-2\omega_0}{\Omega_d} \right)^2 \right]\nonumber\\
\times &\exp&\left[-2\left(\frac{\omega_1-\omega_2}{\Omega_a} \right)^2 \right],
\end{eqnarray}
where $\Omega_a$ ($\Omega_d$) is the bandwidth along the antidiagonal (diagonal) in $\{\omega_1,\omega_2\}$ frequency space.   Note that while $\Omega_a$ may be experimentally controlled through a bandpass filter applied to the photon pairs, $\Omega_d$ may be controlled by the pump bandwidth.

By substituting the transfer function Eq.  (\ref{eq:transfer_function_1layer}) into Eq. (\ref{eq:coincidences}) and define $\tilde{r}_0=r_{01}$, $\tilde{r}_1=r_{12} t_{01} t_{10}$, and $\tilde{r}_2=r_{10} r_{12}^2 t_{01} t_{10}$, we can obtain the following 1-layer interferogram including all echoes and artifacts

\begin{eqnarray}\label{eq:cross_interference}
\frac{C(\tau)}{\Gamma_0} =1&-&\sum _{k=0}^{\infty}V_k \exp\left[-2\left(\frac{\tau-k\, T}{\tau_a}\right)^2\right]\\
&-&\sum _{k,l=0; k<l}^{\infty}V_{kl} \exp\left[-2\left(\frac{\tau-(k+l)\,T/2}{\tau_a}\right)^2\right],\nonumber 
\end{eqnarray}
which represents each of the peaks or dips that appear in the interferogram with visibilities 
\begin{align}\label{eqs:visibilities}
 V_k&=\frac{N_0}{2}\frac{\tilde{r}_k^2}{\Gamma_0},\nonumber\\
 V_{kl}&=\frac{N_0\tilde{r}_k\tilde{r}_l}{\Gamma_0}\exp\left[-\frac{1}{2}\left(\frac{(l-k)T}{\tau_d}\right)^2\right] \\
 & \phantom{-----------}\times\cos\left[\omega_0 (l-k)T\right],\nonumber
\end{align}
and with $\Gamma_0$ expressed as

\begin{align}
\label{eq:self_interference}
\frac{\Gamma_0}{N_0}= \frac{1}{2}\sum_{k=0}^{\infty}\tilde{r}_k^2 &+ \sum _{k,l=0; k<l}^{\infty} \tilde{r}_k\tilde{r}_l\exp\left[-\frac{1}{2}\left(\frac{(l-k)T}{\tau_a}\right)^2\right]&\nonumber\\
 &\times \exp\left[-\frac{1}{2}\left(\frac{(l-k)T}{\tau_d}\right)^2\right]\cos(\omega_0 (l-k)T).&
\end{align}


 In the above equations, $\tau_a\equiv 4/\Omega_a$ and $\tau_d\equiv 4/\Omega_d$ represent the anti-diagonal and diagonal 
widths of the joint spectral intensity.
Fig. \ref{fig:two_boundaries_artifacts} (blue line) shows the simulated QOCT interferogram for a single layer of optical thickness $T/2$ (two interfaces) produced by SPDC photon pairs with a continuous wave (CW) pump. The overlying inset  shows the actual configuration of the sample to identify which dips represent actual interfaces. We consider a sample surrounded by air with reflectivities on each boundary independent of the refractive index contrast between the material and the surrounding air. The number of peaks/dips is consistent with the model given in Eq.~(\ref{eq:cross_interference}), where a set of artifacts appears due to the long two-photon coherence produced by the CW pump laser. 
The QOCT interferogram can be simplified further when
 $T\gg \tau_d$, corresponding to the use of a pulsed pump (e.g. femtosecond-duration) in the SPDC process. Under this condition, $\exp[-(m T/\tau_d)^2/2] \rightarrow 0$ (with  $m\in \mathds{N}$), $V_{kl}\rightarrow 0$, and $\Gamma_0$  reduces to $\Gamma_0 = N_0\left(\sum_{k=0}^{\infty}\tilde{r}_k^2\right)/2$, resulting in \cite{Graciano2019}
\begin{equation}\label{eq:C_reduced}
    \frac{C(\tau)}{\Gamma_0} = 1- \sum _{k=0}^{\infty}V_k \exp\left[-2\left(\frac{\tau-k\, T}{\tau_a}\right)^2\right],
\end{equation}
where $\frac{V_k}{N_0}=\frac{1}{2}\frac{\tilde{r}_k^2}{\Gamma_0}$. Fig. \ref{fig:two_boundaries_artifacts} (red line; vertically-displaced for graphical clarity) confirms this result and shows that for $T\gg \tau_d$, the artifacts disappear in the QOCT interferogram \cite{Yepiz_2019}. The remaining dip labeled by $V_2$ represents an echo.

\begin{figure}
    \centering
    \includegraphics[width=\columnwidth]{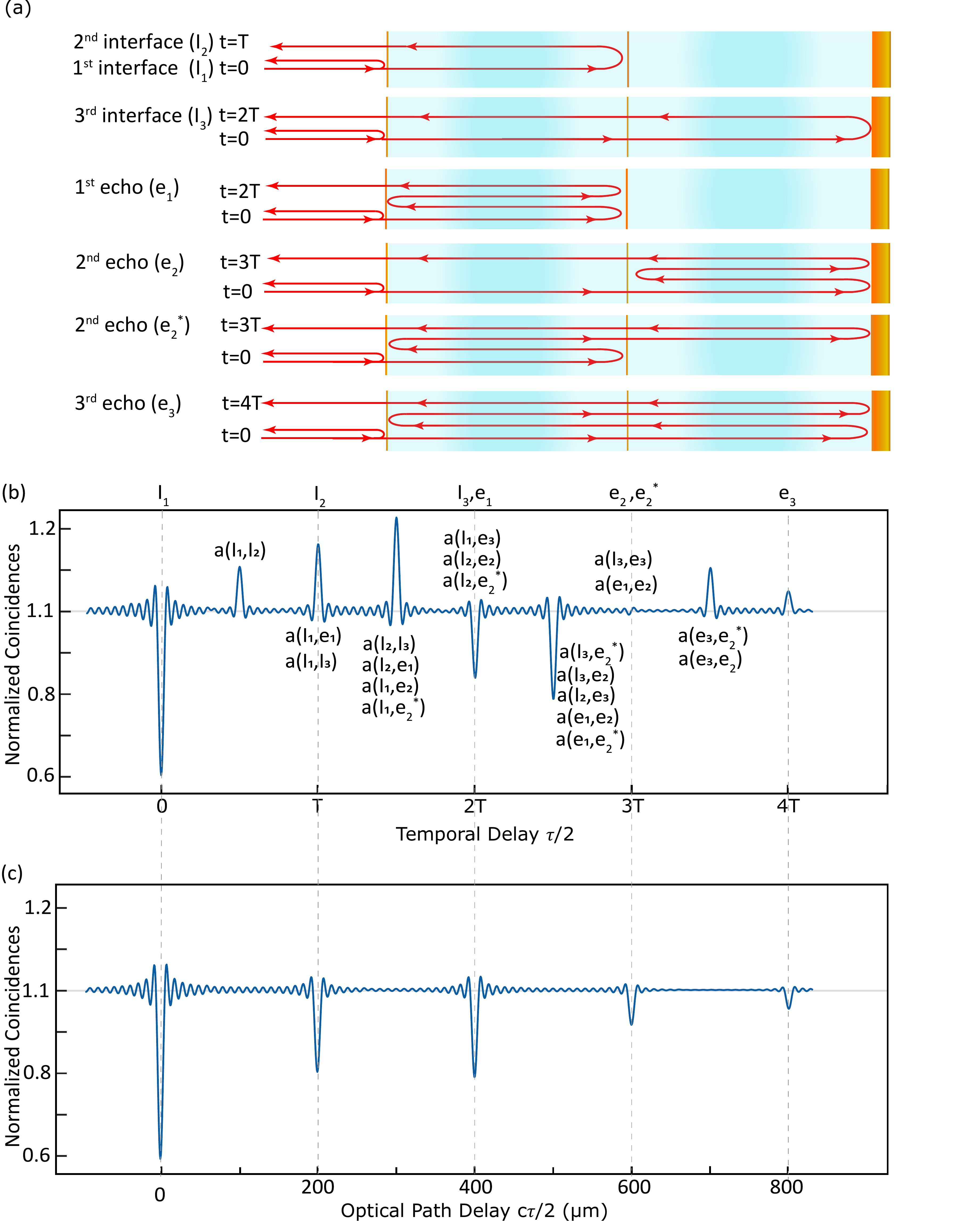}
    \caption{(a) Multiple reflections within a two-layer (three-interface) sample with thickness $d$, refractive index $n$ and with a roundtrip time $T=\frac{2nd}{c}$; we have assumed two layers of identical width of optical width parameter $cT=200\mu$m.   The lowest-order echoes correspond to (see diagram): i) $e_1$ with four passes between interfaces 1 and 2, ii) $e_2$ with two passes between interfaces 1 and 2, and four passes between interfaces 3 and 4, iii) $e_2^*$ with four passes between interfaces 1 and 2, and two passes between interfaces 3 and 4, and iv) $e_3$ with four passes between interfaces 1 and 3.
    (b) Shows the corresponding interferogram using a monochromatic CW laser, where we have indicated which processes amongst HOM dips ($I_j$), echoes ($e_j$), and artifacts ($a(i,j)$ involving interfaces $i$ and $j$) contribute to each peak/dip.
    (c) Shows the interferogram using a pulsed (femtosecond) laser.   Note that sinc-type oscillations arise from a rectangular bandpass filter acting on the signal and idler photons.}
    \label{fig:2layers_peaks_scheme}
\end{figure}

The model presented above  can be used to describe highly-complex QOCT interferograms derived from samples that include multiple layers as well as lossy materials. Fig. \ref{fig:2layers_peaks_scheme} shows the case of a two-layer sample (three interfaces) of equal optical thicknesses $T/2$,  surrounded by air. An equation in closed form such as the one presented for the single-layer case is no longer practical; therefore, we obtain the QOCT trace numerically. The interferogram shows a complex array of peaks and dips derived from interfaces, echoes, and their cross-correlations. The labels in Fig. \ref{fig:2layers_peaks_scheme}(b)  tag all contributions for each peak/dip, in the following way: $I_j$ represents the signal from an actual interface, $e_j$ is the signal from an echo; $a(I_j,I_k)$, $a(I_j,e_k)$, and $a(e_j,e_k)$ are the artifacts coming from the coherence between interfaces $I_j$, actual interfaces with echoes, and between echoes, respectively. This example shows that artifacts and echo-induced artifacts can obscure real interfaces; note that echoes which may themselves be small or imperceptible in a given experimental trace, may lead to major effects on the recovered structure of the sample due to their artifacts.  Finally, as shown in Fig. \ref{fig:2layers_peaks_scheme}(c), the QOCT signal will show echoes due to the multiple reflections that cannot be suppressed through the use of a broadband pulsed SPDC pump.

\subsection{Data processing using a Genetic Algorithm}
\label{sec:GA}

As shown in the previous section, the QOCT interferogram can exhibit a complex structure containing artifacts and echoes that can be difficult to distinguish from the interfaces, which tends to make QOCT impractical for real-world applications. The artifacts can be removed using a pulsed SPDC pump. However, in addition to the cost of using a pulsed laser, this does not prevent the appearance of echoes. Because of this, we tackle the challenge of characterizing the sample using optimization strategies, where the goal is to search for the sample parameters (such as thicknesses and reflectivities) that lead to the best fit with the measurements. Since the complexity of the QOCT interferogram increases with the number of interfaces (see Eq.~(\ref{eq:coincidences})), a searching algorithm for this problem represents a highly nonlinear process for which gradient-based optimization methods are not suitable. For this reason, we have opted to apply a genetic algorithm. 

A genetic algorithm (GA) is an optimization tool based on evolution theory that uses natural selection to solve both constrained and unconstrained optimization problems~\cite{gen, chambers2001}. In a GA, a population of candidate solutions, called individuals, evolves into better solutions mimicking the laws of natural selection.   In the algorithm, the 
properties of a candidate solution are encoded in its ``chromosomes''.  At each iteration of the algorithm, crossover involving pairs of selected individuals (or parents) form the next generation of individuals (or children), while 
chromosomes are also modified through random mutations.  Over many generations, the system parameters can evolve toward a solution.

\begin{figure}[ht]
\centering
 \includegraphics[width=\columnwidth]{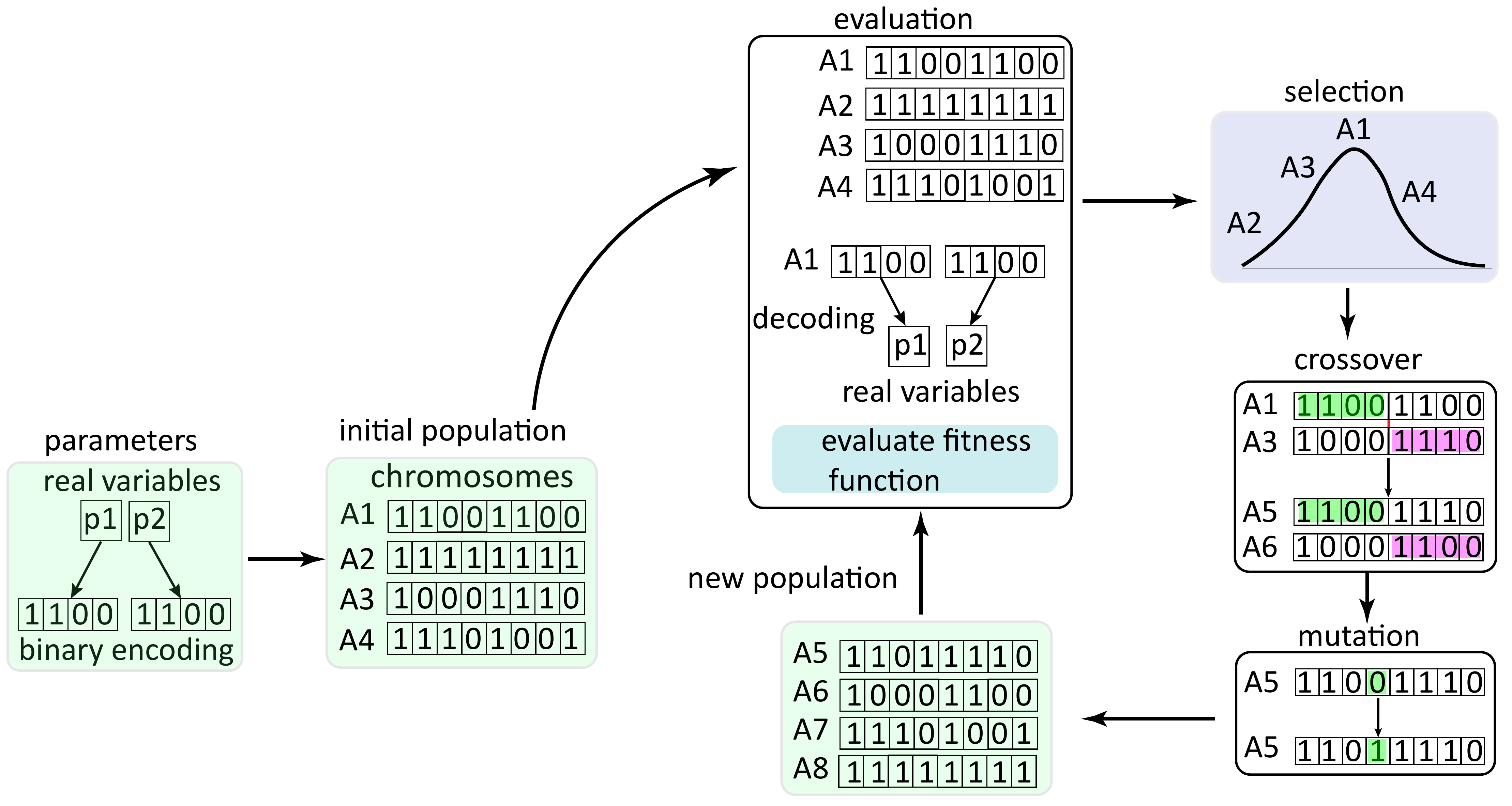}
\caption{\textbf{Genetic algorithm diagram.}  In each iteration of the algorithm, certain solution candidates from the population are selected according to a fitness criterion, which are then  allowed to reproduce. New individuals are obtained by crossover which means combining genetic information from two selected parents.    Mutations are then allowed to occur, which helps to avoid stagnation at local minima.   The algorithm runs until the fitness function reaches a certain threshold value, or until the maximum number of generations is reached. See the main text for further details.
\label{fig:ga_diagram}}
\end{figure}

Figure \ref{fig:ga_diagram} shows the GA stages: initialization, evaluation, selection, crossover, mutation, and replacement of the initial population. The parameters to be optimized are encoded in the  chromosomes that represent each solution candidate. Such chromosomes are randomly chosen so as to obtain an \emph{initial population} solution candidates~\cite{Pawar2015}. At each  iteration of the algorithm, the fitness of each given solution candidate is quantified through a   \emph{fitness function}, which in our case is defined as 
the mean absolute error calculated between the measured QOCT interferogram, $C_\mathrm{exp}(\tau)$, and the retrieved interferogram $C_\mathrm{ret}(\tau; \mathrm{parameters})$, calculated with the parameters being optimized. \emph{Selection} allows us to choose, from the population, the individuals considered as the aptest, which are allowed to reproduce.
For example, in Figure~ \ref{fig:ga_diagram}, from the available set of individuals \{A1, A2, A3, A4\}, only individuals A1 and A3 are selected for the crossover.  \emph{Crossover} consists of the combination of the genetic material of two sets of chromosomes (the parents)  for example, the first half of A1's genes (marked in green), are combined with the second half of A3's genes to produce a new individual A5, or switching which half is taken from each parent leads to an alternative new individual A6.
In the \emph{mutation} stage, a fraction of the genes in a given solution candidate is subjected  to a random change or mutation;  this provides genetic variability, which helps to avoid stagnation at local minima.

Our algorithm was scripted in Matlab, using the Global Optimization toolbox which takes advantage of parallel computation. In our case, the parameters of interest are the intra-layer distances, loss parameters, and interface reflectivities.  For a sample with $N$ interfaces, we have $4N-3$ effective real-valued parameters where for each of the $N-1$ transmissive segments we have one optical distance and one loss parameter. This results in $2(N-1)$ effective parameters and thus yielding a total $4N-3+2(N-1)=6N-5$ effective real-valued parameters (see the Supplemental Material).
When running the GA one selects some parameters to be given fixed values \emph{a priori}, while the rest are concatenated in a binary chain forming the chromosome. 
We use a population size of 300 individuals, with around 50 generations, depending on the complexity of the problem. Convergence to a solution is typically achieved in a time between 3 and 10 minutes.

\section{Experiment}
\subsection{Experimental setup and sample preparation}
\begin{figure*}[ht]
\centering
\includegraphics[width=\linewidth, scale=0.8]{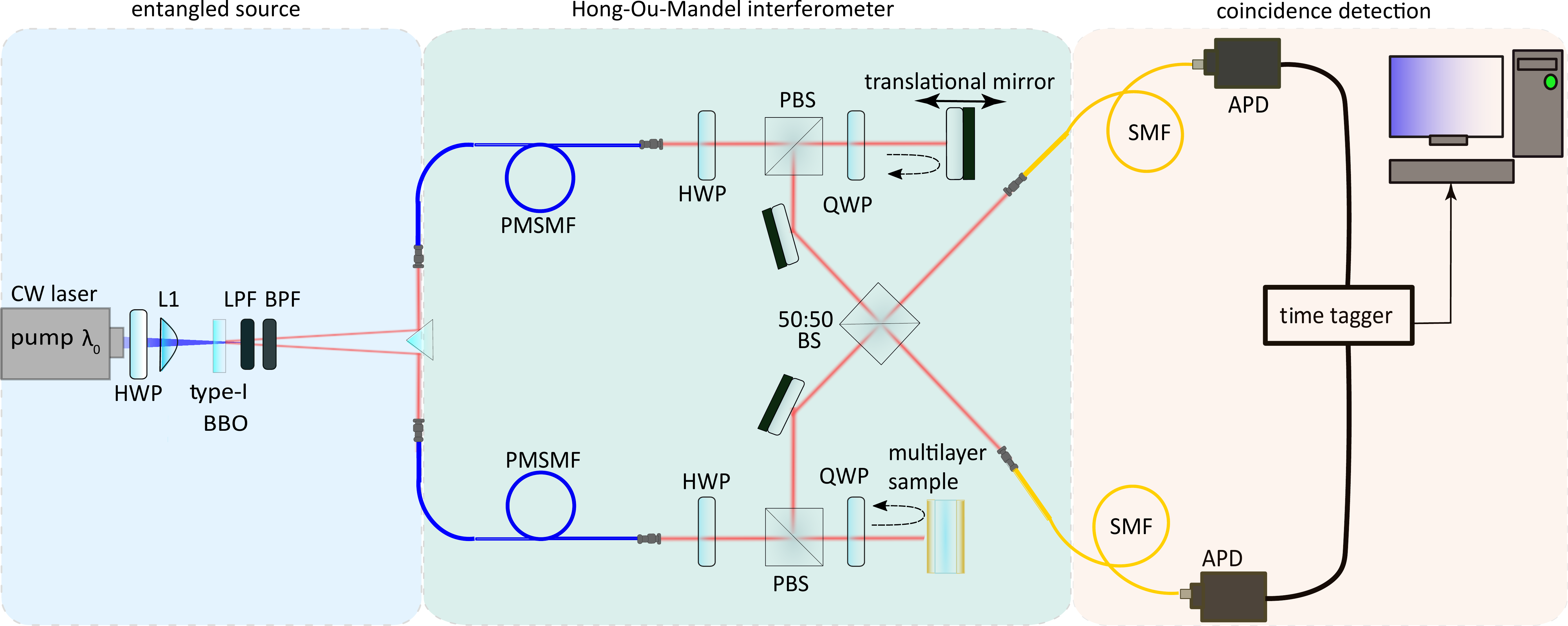}
\caption{Quantum Optical Coherence Tomography setup consisting of the  Spontaneous Parametric Down Conversion entangled photon pair source, Hong Ou Mandel (HOM) interferometer, and coincidence detection system. In the entangled source, a tunable continuous-wave (CW) laser with wavelength $\lambda_{0}=404.5$ nm is used to pump type-I $\beta$-Barium Borate (BBO) nonlinear crystal. The output of the crystal is then fed into the HOM interferometer. The idler-signal pairs are then recombined and measured using avalanche photodiodes (APD).  Here single-mode fibers (PMSMFs), polarization beam splitter (PBS), quarter waveplate (QWP), beam splitter (BS), low/band-pass filter (L/BPF), single-mode fiber (SMF), and half-wave plate (HWP).
}
\label{fig:QOCT_setup2}
\end{figure*}

Our experimental setup consists of three stages: the entangled photon pair source, the HOM interferometer, and the coincidence detection system, see Figure~\ref{fig:QOCT_setup2}.

In the source, a tunable continuous-wave (CW) laser (\textit{Moglabs LDL}) is used to pump a 2-mm long $\beta$-Barium Borate (BBO) nonlinear crystal.
The laser is initially centered at $\lambda_{0}=404.5$ nm and set to a power of 60 mW. Its output polarization is controlled by a half-wave plate (HWP) which is followed by a plano-convex lens (L1, $f=1000$ mm) that focuses the beam on the crystal plane to a waist of 300 $\mu$m diameter. The BBO crystal is cut at $29.2^{\circ}/90^{\circ}$ for type-I SPDC phase matching that allows non-collinear generation of entangled single photon pairs with an exit angle of $3^\circ$ with respect to the pump axis. At the output of the BBO crystal the pump light is filtered by a long-pass filter (LPF) with cut-off wavelength $\lambda = 500$ nm (\textit{Thorlabs} FELH0500) and the stream of photon pairs is spectrally filtered by a 40 nm band-pass filter (BPF) centered at 800 nm (\textit{Thorlabs} FBH800-40). The signal and idler photons are then further separated by a right-angle prism mirror and coupled into polarization-maintaining single-mode fibers (PMSMFs) in the reference and sample arms, respectively. This allows photon spatial mode filtering and ensures the propagation of a Gaussian mode in each of the interferometer arms.


After exiting the single-mode fiber, the signal photon feeds the reference arm where it traverses a temporal delay system composed of a polarization beam splitter (PBS), quarter waveplate (QWP), and a translatable mirror. Linearly polarized light is transmitted through the polarization beamsplitter and reflected back from the mirror. As it passes twice through the quarter waveplate, the outgoing polarization becomes orthogonal with respect to the input and is now reflected by the polarization beamsplitter.  The mirror is mounted on a motorized precision linear stage (Newport MFA-CC) that permits control of the temporal delay between signal and idler photons $\tau$ through a path length difference of the reference arm relative to the fixed-length sample arm.

The idler photon, on the other hand, feeds the sample arm with a similar path in which 
a stationary sample composed of multiple reflective interfaces is positioned in place of the translatable mirror of the reference arm.  
The two photons, signal and idler, are recombined at the 50:50 beamsplitter (BS) where the HOM interference occurs. The modes corresponding to the two beamsplitter outputs are coupled into avalanche photodiodes ($\mathrm{APD}$; \textit{Perkin Elmer} SPCM-AQR-14-FC) and the HOM interferogram is acquired by a time controller unit (IDQ-ID900) based on the coincidence rate $C(\tau)$, with a 2 ns coincidence window. 

The coincidence interferograms in our experiment are obtained by translating the reference mirror in 1 $\mu m$ steps and recording the coincidence rate at each position. Since every sample interface involves a different sample arm length, the interferometer is balanced at multiple points. This is achieved by compensating the additional sample arm length with the variable reference arm length. As a result, a HOM dip appears in the interferogram for each interface and the distance between such dips is proportional to the separation between real interfaces.

The samples in our experiment consist of one or two  glass substrates (index of refraction $n=1.52$ at 800 nm) coated with gold thin films through electron beam deposition.
The two-interface sample is a $\sim180\,\mu m$-thick glass substrate with coatings of  thicknesses 8 nm and 50 nm that form the front and back surfaces, respectively,  corresponding to nominal intensity reflectivities $R_{01}=0.36$, $R_{10}=0.34$ (air-glass), and $R_{12}=0.95$ (glass-air).
The three-interface sample consists of two independent $\sim200\,\mu m$ glass substrates coated in a similar fashion and pressed together. The first glass substrate is coated with 10 nm and 5 nm thick gold films while the second substrate is 
uncoated on one side and coated with a 50 nm  film on the other side;  the uncoated side is pressed against the 5 nm coating of the first substrate. The 5 nm, 10 nm, and 50 nm gold films form respectively the front, middle and back surfaces of the sample.  Note that since the glass substrates do not adjoin perfectly, an air gap between them is present. The presence of the air gap implies that the sample is well described by four interfaces, with nominal intensity reflectivities: i) $R_{01}=0.46$ and $R_{10}=0.44$, for air-first coverslip, ii) $R_{12}=0.22$ and $R_{21}=0.20$, for first coverslip-air, iii) $R_{23}=0.04$ and $R_{32}=0.04$, for air-second coverslip,  and iv) $R_{34}=0.95$ for second coverslip-air.

The gold coating thicknesses are estimated from the settings used during the electron beam deposition procedure and are within a tolerance of 0.5 nm. The associated error of the reflectivities are calculated using the Fresnel equations and thin-film theory \cite{Heavens_1960, Mohammed_2019} and is less than 4.3\%.


\subsection{Results and discussion}

\begin{figure}
    \centering
    \includegraphics[width=\columnwidth]{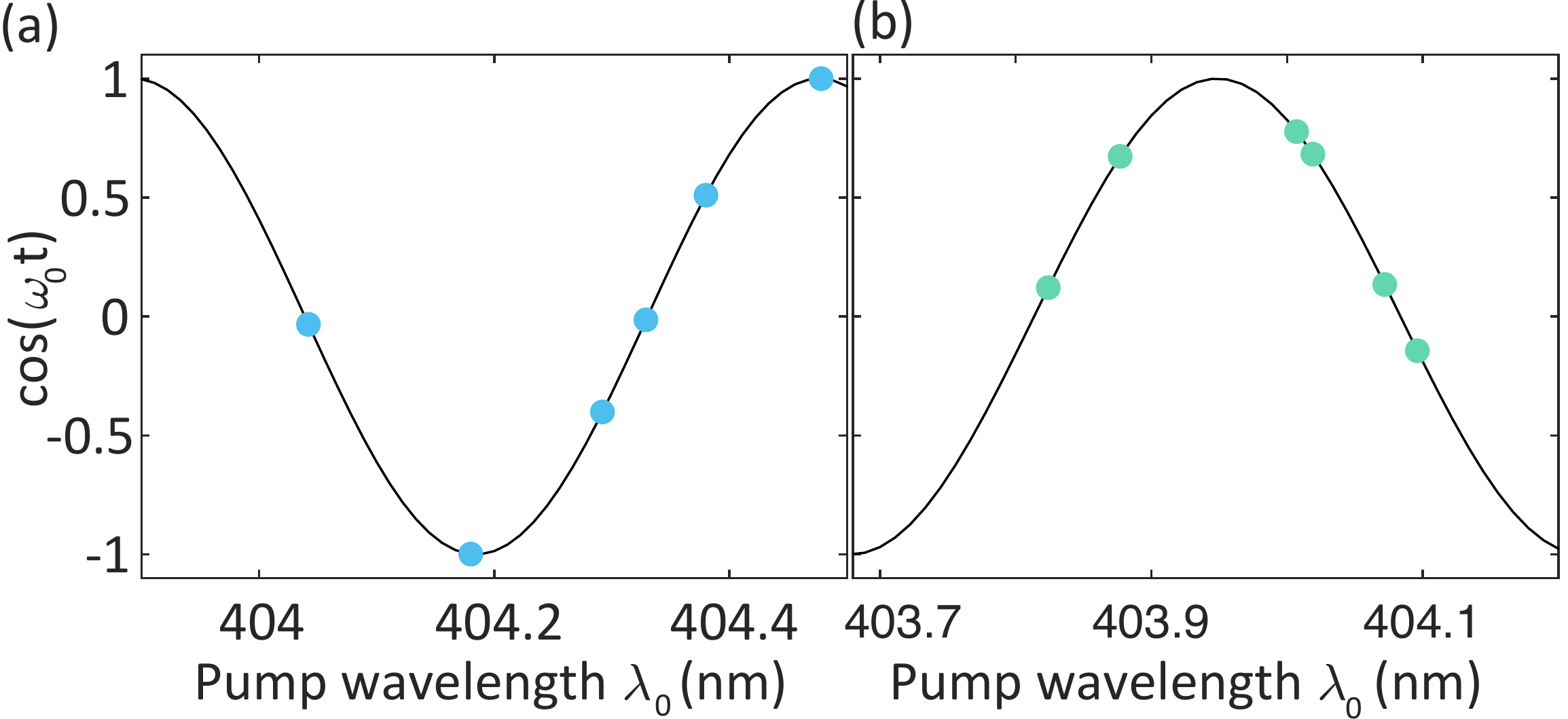}
  \caption{Cosine functions ruling the visibility of the cross-interference artifact between the front and back surfaces,  for a two-layer (a) and three-layer (b) sample.  The dots indicate the pump wavelengths used in the experiment.}
  \label{fig:cosines2}
\end{figure}

It is of interest to illustrate how our approach operates in the context of the particular peak/dip composition exhibited by the QOCT interferogram. 
Note that our measurements are realized by tuning the pump laser wavelength $\lambda_{0}$ to particular values, selected to yield a peak, dip, or no artifact as predicted from the sign and amplitude of the $\mathrm{cos}(m \omega_{0} T)$ function (with $m=1,2$) in the visibility parameters Eq. (\ref{eqs:visibilities}). The argument of the cosine function depends on the time $T$ taken by the signal photon to  travel between the two interfaces responsible for creating the artifact; thus, the  cosine function period depends on the distance between them. Since each artifact is created by two features (interface-interface, interface-echo, echo-echo) with a well-defined distance between them, it has its own cosine function that defines its shape for a given pump wavelength.
This means that we can target a specific artifact by choosing the corresponding cosine function for pump wavelength tuning, see Fig. \ref{fig:cosines2}. We have used such an approach to generate a set of interferograms with different shapes for the same sample. As a result, we are able to test whether for a given sample our algorithm gives the same parameters for different input wavelengths, as the underlying sample morphology is the same.

\begin{figure*}
    \centering
    \includegraphics[width=\textwidth]{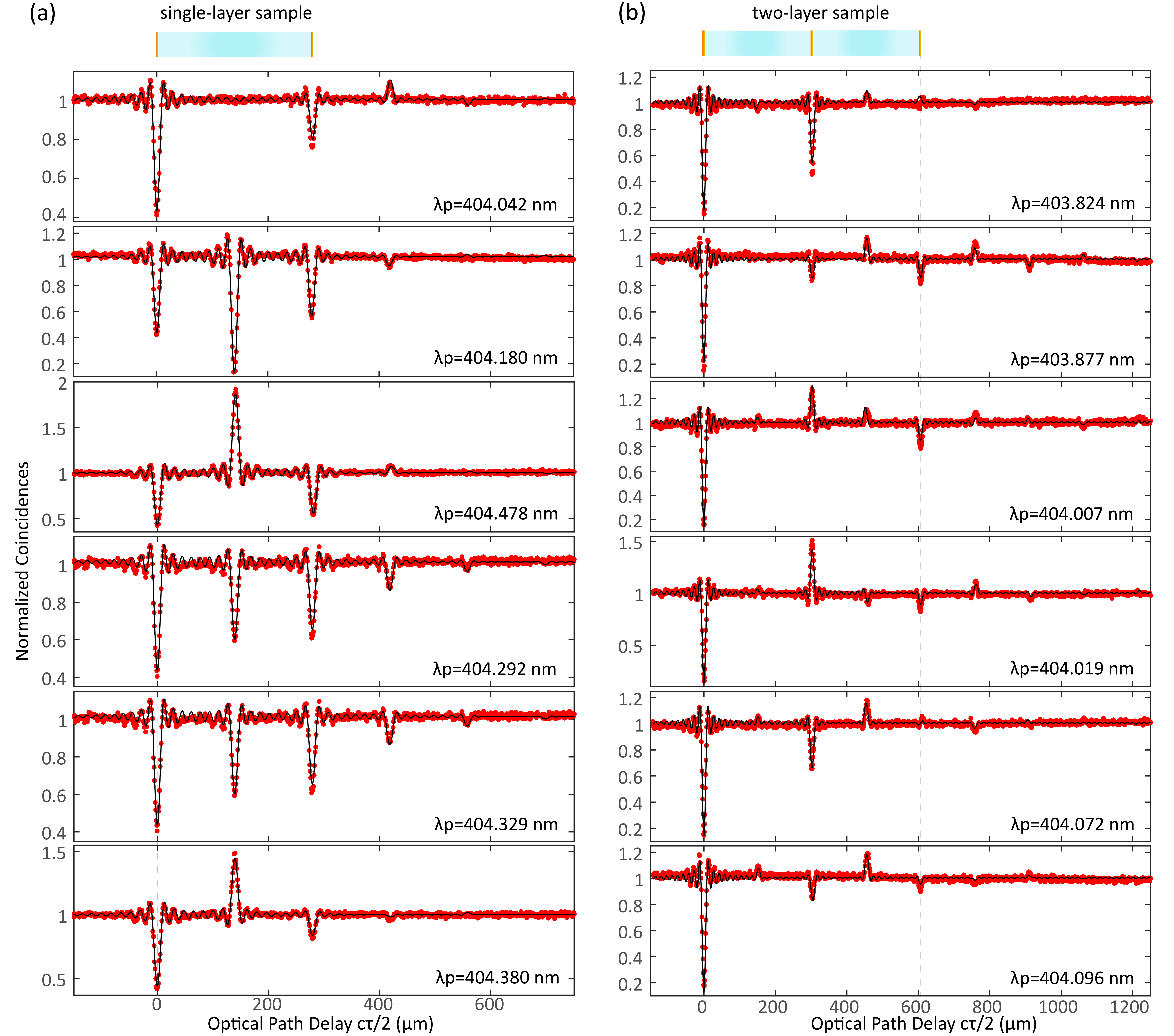}
    \caption{Comparison between the experimental data (red dots) and its corresponding fitting curve (black solid line) as optimized by the genetic algorithm. (a) Shows the results for the single layer (two interface) sample, whereas (b) shows the comparison for the two-layer sample. For reference, the front surface of the sample is set to $\tau = 0$}
    \label{fig:stack1&3layers_main}
\end{figure*}

Fig. \ref{fig:stack1&3layers_main}(a) shows the resulting interferograms for a two-interface sample upon changing the target artifact created by the front and back surfaces of the sample. For example, the zero value of the cosine function at $\lambda = 404.042$ nm yields an artifact with zero amplitude, while a minimum in the cosine function at $\lambda = 404.180$ nm results in the  dip with the largest visibility possible, and so on. 

For a three-interface sample, the complexity of the signal increases, see Fig. \ref{fig:stack1&3layers_main}(b). 
There are more artifacts present in the interferogram. For this case, we decided to target also the artifact associated with the front and back surfaces. The situation is more complex than for the single-layer case because the artifact overlaps with the dip corresponding to the middle interface. While the corresponding cosine function predicts a suppressed artifact for 404.096 nm, the interferogram for that wavelength is not flat even though the artifact amplitude is zero. The net dip comes from the contribution of the middle interface. 

The Genetic Algorithm was programmed to estimate the width of each layer and the reflectivities of the corresponding interfaces by increasing, in each epoch, the number of layers $N$ that the unknown sample could have. For each guess of $N$, the fitness value was retained and used as a figure of merit for the search procedure. The GA is initially given a guess $R_{0}$ and starts by assuming that the sample is single-layered, $N=1$, and the optimization calculates a fitness value for this case. Next, the GA assumes a two-layer sample; i.e. $N=2$, and calculates its fitness value. The procedure is repeated with increasing values of $N$ until a minimal fitness value is obtained. The best set of parameters corresponds to the case of $N_{\mathrm{best}}$ layers for which the reconstructed interferogram best matches the experimental signal. Note that at the step of this implementation in which we assume $N$ interfaces, we set the values of the first $N-1$ reflectivities.  While this is meant as a proof-of-principle demonstration, the possibility of for example only setting the values of the reflectivities which could be amenable to direct experimental determination is left for future analysis (see also the discussion in the Supplemental Material).  As previously stated, in each step of the searching GA the QOCT interferogram is evaluated from the fitting parameters, which assumes a perfect monochromatic pump beam, i.e. with a zero SPDC pump spectral width, and that the generated photon pairs are centered around 808 nm with a width of 40 nm.

Fig.~\ref{fig:stack1&3layers_main} shows how the reconstructed interferogram matches the experimental data after the GA has ended its search. The results for the single-layer sample are presented in Fig.~\ref{fig:stack1&3layers_main} (a) while the results for the two-layer sample are shown in Fig.~\ref{fig:stack1&3layers_main} (b). As seen from the plot, the reconstructed interferograms are in excellent agreement with the experimental traces, for both cases.  In particular, for the case of the single-layer sample, the artifact can be clearly identified by changing the central wavelength that causes it to flip from a dip to a peak. 
This behavior, however, becomes problematic for a sample with a higher number of layers. For instance, in the case of the two-layer sample, the artifact produced by the front and back surfaces  
indeed changes from a dip to a peak when the pump wavelength is changed. However, it overlaps with the signal produced by the middle surface. 
As a consequence, simply filtering the signals that change their behavior as the pump wavelength is varied could potentially remove structures associated with real interfaces. 

\begin{table}[t]
    \centering
    \begin{tabular}{ccc}
    \toprule
 Wavelength (nm) & $d_1$ ($\mu$m)  & $R_1=|r_1|^2$  \\ \midrule
 404.036 & 185.227 & 0.896 \\
 404.042 & 184.566 & 0.916 \\
 404.180 & 184.801 & 0.876 \\
 404.292 & 184.826 & 0.891 \\
 404.329 & 184.705 & 0.946 \\
 404.380 & 185.091 & 0.942 \\
 404.478 & 186.278 & 0.883 \\ \midrule
  & 185.213 & 0.907\\ \bottomrule
    \end{tabular}
    \caption{Optimal parameters for a one-layer sample with fixed parameters: $R_{01}=0.310$, $R_{10}=0.290$, and metal thickness $\ell_{au}=7.5$ nm,  for different pump wavelengths. 
    The last row is the average of the retrieved widths and reflectivities over all realizations.
    }
    \label{tab:optParam_1layer}
\end{table}

\begin{table}[t]
    \centering
    \begin{tabular}{ccccc}
    \toprule
    Wavel. (nm) & $d_1$ ($\mu$m) & $d_2$ ($\mu$m) & $d_3$ ($\mu$m)  &
   $R_4$ \\ \midrule
   403.824 & 199.162 & 1.844 & 202.453 & 0.933  \\
    403.877 & 199.118 & 0.064 & 203.334 & 0.879  \\
    404.007 & 198.093 & 3.288 & 200.960 & 0.907  \\
    404.019 & 197.494 & 1.302 & 203.058 & 0.974  \\
    404.072 & 200.731 & 1.985 & 199.886 & 0.889  \\
    404.096 & 202.990 & 2.171 & 196.822 & 0.899  \\ \midrule
    ~ & 199.598 & 1.775 & 201.086 & 0.914  \\
    \bottomrule
    \end{tabular}
    \caption{Optimal parameters retrieved, for different pump wavelengths, for our two-layer sample with the following fixed parameters $R_{01}=0.460$, $R_{10}=0.440$, $\ell_{au,0}=10$ nm (first metal layer),  $R_{12}=0.202$, $R_{21}=0.190$, $\ell_{au,1}=5$ nm (second metal layer),  $R_{34}=0.04$, and $R_{43}=0.04$. The last row is the average of the retrieved features over all realizations.}
    \label{tab:optParam_3layer}
\end{table}

The optimal parameters found for both the single- and two-layer samples are summarized in Tables~\ref{tab:optParam_1layer} and \ref{tab:optParam_3layer}. Note that because the two-layer sample is built from two coverslips pressed together, our GA  treats it as a four-interface sample, with an air gap between the second and third interfaces, leading to three distances and four reflectivities.   We have provided the algorithm with the value of the  reflectivity  of the first interface (in the case of the 1-layer sample), and the reflectivities for the first three interfaces (in the case of the 2-layer sample), so that in both cases the algorithm is left to find a value for the last reflectivity.  The algorithm predicts well the distances, although there is a larger fluctuation in the case of the air gap thickness. Additionally, we note that the reflectivity values of the last interface predicted by the algorithm are in all cases reasonably close to the nominal values. We ascribe the variations observed in the recovered parameters to the presence of noise or fluctuations in the system.  For an additional discussion on the determination of the morphology of more complex samples and the effect of noise, please see the Supplemental Material.

Note that the width of our HOM dips is around $11\mu$m which is significantly wider than the width of the airgap $\approx 2-3\mu$m.	While we cannot resolve two dips separated by such a short distance, the airgap has an effect on the reflectivities of the second and third interfaces (adjacent to the airgap), which in turn affects the dip and artifact visibilities of the interferogram as a whole.   In this manner, the airgap may have an effect on the overall interferogram beyond the presence of two nearly-coincident dips which appear as a single dip in the experimental interferograms.

We point out that the experimental results presented in this paper benefit from the usual factor of $2$ improvement in axial resolution for QOCT as compared to an equivalent classical apparatus (OCT) \cite{Abouraddy_2002}. In addition, it is known that for the case of a two-photon state which exhibits strict spectral anti-correlations, even-order (in particular quadratic) dispersion effects are suppressed.  Such spectral anti-correlation is ensured by the use of a narrow-linewidth continuous wave laser as SPDC pump, which is indeed the case in our experiment (the linewidth of our pump laser is $200$kHz).   Thus, our experimental results benefit from quadratic dispersion cancellation.

\section{Conclusions} 

We have presented a new, more complete model for Quantum Optical Coherence Tomography (QOCT) which accounts for the effects of 
real interfaces, artifacts, echoes, as well as the superposition of all these.  A thorough understanding of these structures can give richer information beyond the separations between the inner interfaces within a multilayer sample. We have changed the paradigm and 
transitioned from suppressing artifacts to inferring the source morphology giving rise to a given interferogram,  fully taking into account all  artifacts and multiple reflections.
Solely removing peaks or dips from the interferogram can be a naive approach as the morphology of the sample is generally unknown and artifact removal can be counterproductive since the technique comes with the risk of information loss. 

Our new model exposes the complexity of the QOCT interferogram arising from the quantum nature of the source, which is affected by multiple reflections within the sample. While artifacts have been previously identified, echoes and echo-induced artifacts, as well as their possible superposition on other peaks or dips had not been reported.  We have experimentally demonstrated that a given artifact can be effectively removed from the signal by tuning the pump frequency and consequently shaping its visibility to extinction. Echoes, on the other hand, survive the process and contribute to the appearance of intricate structures in the interferogram.
The importance of our genetic algorithm thus lies in its ability to identify with plausibility the number of layers in a sample, as well as intra-interface distances and interface reflectivities.
With further upgrades, our approach presented here could bring the possibility of monitoring small-scale light-sensitive biological entities such as living cells and bacteria without introducing optical damage.

\subsection*{Acknowledgments}
S.B. acknowledges funding by the Natural Sciences and Engineering Research Council of Canada (NSERC) through its Discovery Grant, funding and advisory support provided by Alberta Innovates through the Accelerating Innovations into CarE (AICE) -- Concepts Program, and support from Alberta Innovates and NSERC through Advance Grant. This project is funded [in part] by the Government of Canada. Ce projet est financé [en partie] par le gouvernement du Canada. AU acknowledges Consejo Nacional de Ciencia y Tecnolog\'{i}a (CF-2019-217559),  PAPIIT-UNAM (IN103521), and AFOSR (FA9550-21-1-0147). VS acknowledges the support from Alberta Innovates grant. 





\bibliographystyle{ieeetr}

\newpage
\onecolumngrid
\begin{center}
\textbf{\large Supplementary Materials}
\end{center}

In this supplemental material section, we discuss the power of our genetic algorithm for the determination of the sample morphology, particularly providing an example of a more complex five-interface sample.  This example involves an interferogram computed from our model (rather than obtained experimentally) so that there is a complete absence of noise.
\begin{center}
\textbf{An illustrative morphology determination example}
\end{center}

We consider a sample with $N$ dielectric interfaces each of which is assumed  to be well-described as lossless, and we separately consider the effects of a metallic layer through a loss matrix.  Each interface is described by two complex-valued reflectivities (forward and backward) as well as  two complex-valued transmissivities (forward and backward).  This gives 4 complex-valued parameters  or 8 real-valued parameters per interface.   Now, for a lossless interface (between media $l$ and $k$) there are 4 constraints on these 8 parameters: $|r_{kl}|=|r_{lk}|$,  $|t_{kl}|=|t_{lk}|$,  $|t_{kl}|^2+|r_{kl}|^2=1$,  and $t_{lk}/t_{kl}*=- r_{lk}/r_{kl}*$.    This leaves 4 effective real-valued parameters per interface. For the last interface, it is only the forward reflectivity that can have an effect on the QOCT interferogram, so this last interface is described by a single effective parameter.   For the $N$ interfaces, we then have $4N-3$ effective, real-valued parameters.  In addition, for each of the $N-1$ transmissive segments (corresponding to $N$ interfaces) we have one optical distance and one loss parameter, yielding $2(N-1)$ effective parameters. In total, for $N$ interfaces we then have $4N-3+2(N-1)=6N-5$ effective real-valued parameters. Thus, the complexity of the problem could be considered to scale with $N$ as $6N-5$.

 \begin{figure*}[ht]
\label{fig:SM}
    \centering
    \includegraphics[width=\linewidth, scale=1.2]{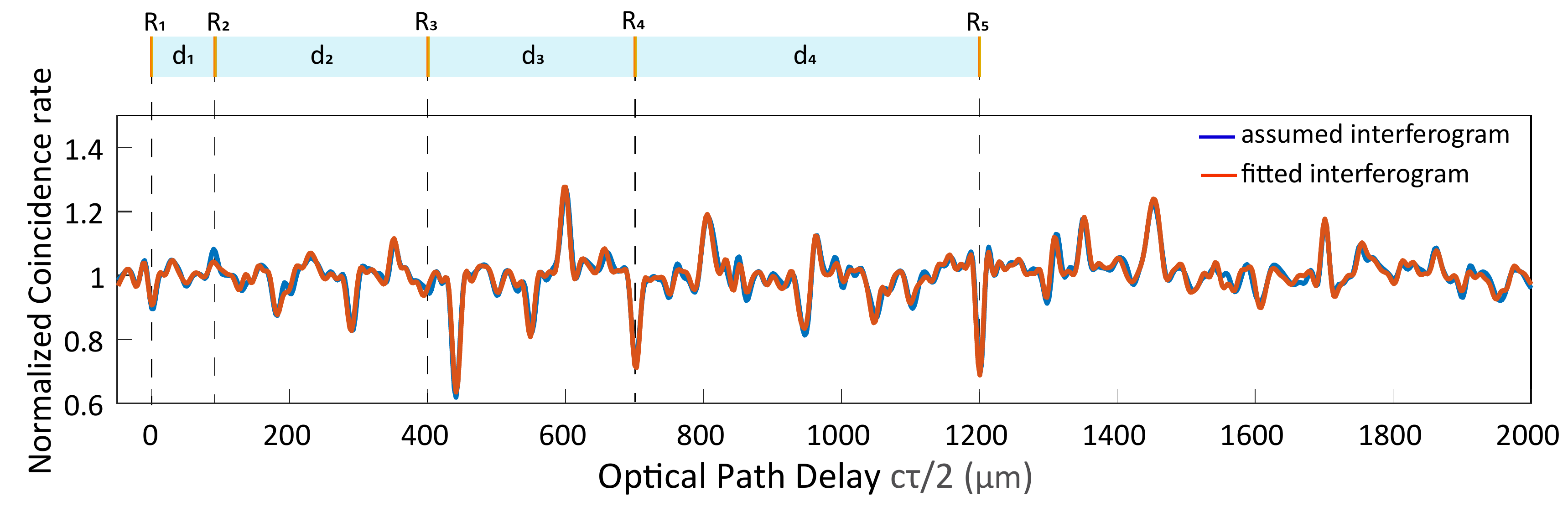}
    \caption{QOCT interferogram obtained from the model for a particular four-layer (five-interface) sample (blue line), together with the fitted interferogram derived from our GA algorithm (red line).   The sample morphology is indicated in the overlayed inset.}
    \label{fig:four_layer_fit}
\end{figure*}

In principle, the problem of determining the sample morphology from the interferogram corresponds to the determination of these $6N-5$ real-valued parameters from the interferogram.   The visibility of each of the features (dip or peak) in the interferogram provides a constraint (an equation) so that if the number of equations can be made equal to the number of parameters, we may be in a position to extract the values of all parameters. Let us consider as an example the case of 2 interfaces (1 layer).   In this case, in the interferogram we may observe features corresponding to: i) the two dips, ii) the artifact between the two interfaces, and iii) the echoes as well as artifacts involving echoes. Some of these features involve multiple contributing processes as described in the main text, for example, the second dip is overlapped with the artifact between the first interface and the first echo. If we consider the two dips and the first artifact, that only gives us 3 constraints, while the number of effective parameters is $6\times 2-5=7$.  If we consider in addition the first echo as well as the artifact between the second interface and the first echo, that gives us 5 constraints, still short of the 7 parameters. In the absence of noise so that we may reliably observe the second echo as well as the artifact between the first two echoes, that gives us the 7 constraints that we need in order to extract all 7 parameters. The optimization carried out by our GA algorithm can be thought of as being equivalent to solving a number of coupled equations derived from each of the interferogram features as described above.  Note also that actually solving such a system of equations for an unknown sample may not be practical, as it would hinge on being able to fully identify each of the interferogram features which as will be shown below for a specific example can become extremely challenging.   

\begin{table}[ht]
\label{tab:SM}
    \centering
    \begin{tabular}{ccc}
    \toprule
  & Nominal  & Retrieved  \\ \midrule
 $s_1$ ($\mu$m) & 90 & 91.116 \\
 $s_2$ ($\mu$m) & 110 & 107.440 \\
 $s_3$ ($\mu$m) & 150 & 152.100 \\
 $s_4$ ($\mu$m) & 250 & 249.730 \\ \midrule
 $R_1=\vert r_2\vert^2$ & 0.1 & -  \\
 $R_2=\vert r_2\vert^2$ & 0.1 & 0.096 \\
 $R_3=\vert r_3\vert^2$ & 0.1 & 0.095 \\
 $R_4=\vert r_4\vert^2$ & 0.5 & 0.506 \\
 $R_5=\vert r_5\vert^2$ & 0.9 & 0.901 \\ \bottomrule
    \end{tabular}
    \caption{Retrieved vs nominal optical distance $s_j$ ($j=1,2,3,4$)  and (intensity) reflectivity parameters $R_j$ ($j=1,2,3,4,5$), for a particular five-interface sample.  Note that we have provided the algorithm with the correct value of the first reflectivity (0.1), leaving the other 4 reflectivities as free-fitting parameters.
    }
    \label{tab:optParam_1layer}
\end{table}

In an actual experiment with noise and other experimental imperfections, it becomes rapidly difficult to resolve progressively smaller features from the noise. We may conclude that while for an ideal (noiseless) situation we could be in a position to unambiguously determine all effective parameters from the interferogram, in an actual experiment this will most likely not be the case. To illustrate this discussion, we present in Figure \ref{fig:SM} an interferogram (blue line) \emph{calculated} using our QOCT model for a particular five-interface losssless sample with assumed (nominal) parameter values shown in table \ref{tab:SM}.  Note that for such a complex sample it becomes difficult to identify particular features in the interferogram, a challenge which becomes more acute as the number of interfaces is increased. We now take this interferogram as input for our algorithm, which then outputs both a fitted interferogram (red line) and  provides values for the retrieved parameters.   It is notable that in this case, because there is a complete absence of noise (the interferogram is derived from our model and not from the experiment), the quality of the fit is excellent and the retrieved parameters match closely the nominal ones.  This highlights the power of our GA approach for the determination of the morphology.
\end{document}